# EXISTENCE AND REGULARITY OF WEAK SOLUTIONS OF DEGENERATE PARABOLIC PDE MODELS FOR THE PRICING OF SECURITY DERIVATIVES

RASOUL BEHBOUDI* AND YOU-LAN ZHU†

**Abstract.** This work is focused on the solvability of initial-boundary value problems for degenerate parabolic partial differential equations that arise in the pricing of Asian options, and on the investigation of differential and certain qualitative properties of solutions of such equations. The generalized solvability for such models with degeneracy at the boundaries is proved by employing solutions obtained from finite difference numerical schemes. Furthermore, the regularity of such solutions is studied.

## Introduction:

The price of a discretely sampled Asian option is modeled by a final-value parabolic partial differential equation on an infinite domain that takes the form ([14]):

$$\frac{\partial W}{\partial t}+\frac{1}{2}\sigma^2\eta^2\frac{\partial^2 W}{\partial \eta^2}+\left(D_0-r\right)\eta\frac{\partial W}{\partial \eta}-D_0W=0, \quad -\infty<\eta<\infty \ ,$$

along with a final condition at expiry *T*. In this model, $\eta=\dfrac{D}{S}$, $W(\eta,t)=\dfrac{V}{S}$, $V=V(S,t)$ is the price of option, *S* is the value of the underlying asset, *t* is the time, $D_0$ is the dividend rate, $\sigma(S,t)$ is the volatility, $r(t)$ is the interest rate, and *D* is a constant related to the double average Asian option.

From the numerical point of view, the above problem has to be solved on an artificial finite domain by adding artificial boundary conditions. However, such boundary conditions may not be appropriate in some cases as they would contribute significantly to the error of the numerical work.

By introducing new variables, the above formulation can be transformed into an initial-boundary value problem on a finite domain as follows ([14]):

* Department of Mathematics, University of Phoenix, Charlotte, NC 28273, U.S.A. (rbehboudi@email.phoenix.edu)

† Department of Mathematics, University of NC at Charlotte, Charlotte, NC 28223, U.S.A. (yzhu@uncc.edu)

$$\frac{\partial U}{\partial \tau}=\frac{1}{2}\sigma^2\xi^2\left(1-\xi^2\right)\frac{\partial^2 U}{\partial \xi^2}+\left(D_0-r\right)\xi\left(1-\xi\right)\frac{\partial U}{\partial \xi}-\left(D_0\left(1-\xi\right)+r\xi\right)U, \tag{1}$$

$$0 \le \xi \le 1,\ \tau > 0, \tag{2}$$

$$U(\xi,0)=f(\xi). \tag{3}$$

Since this new equation degenerates at the two boundaries, the solution values at the boundaries can be determined exactly by the ordinary differential equations and initial conditions governing the boundaries. Consequently, in the process of obtaining numerical solutions, no artificial boundary conditions are needed.

In order to justify the validity of this approach, the existence and the uniqueness of solutions of such parabolic equations have to be investigated. Note that since the diffusion coefficient is zero at the boundaries, the usual elliptic arguments do not apply. In particular, the diffusion operator is not strongly elliptic and its graph norm is not equivalent to the $W_2^2$ Sobolev space norm, hence the existence of the solution is not trivial. On the other hand, once the existence of the solution has been established, the uniqueness property would quickly follow from Gronwall inequality.

## ANALYTIC INVESTIGATION OF THE UNIQUENESS OF SOLUTION:

Degenerate parabolic formulations such as (1)-(3) can be written in the following general conservative form:

$$\frac{\partial u}{\partial t}=\frac{\partial}{\partial x}\left(A(x,t)\frac{\partial u}{\partial x}\right)+B(x,t)\frac{\partial u}{\partial x}+C(x,t)u, \tag{4}$$

$$0\le x\le 1,\ t>0, \tag{5}$$

$$u(x,0)=f(x), \tag{6}$$

$$A(x,t)\ge 0,\ A(0,t)=A(1,t)=0, \tag{7}$$

and with one of the following boundary characteristics for $B(x,t)$:

$B(0,t)\ge 0,\ B(1,t)\le 0$ (8.a)
$B(0,t)\ge 0,\ B(1,t)> 0$ (8.b)
$B(0,t)< 0,\ B(1,t)\le 0$ (8.c)
$B(0,t)< 0,\ B(1,t)> 0$ (8.d)

For simplicity, we assume that $C(x,t)=0$. Then multiplying both sides of (4) by $2u$, we obtain:

$$\frac{d}{dt}\left(u^2\right)=2\frac{\partial}{\partial x}\left(Au\frac{\partial u}{\partial x}\right)+B\frac{\partial}{\partial x}\left(u^2\right)-2A\left(\frac{\partial u}{\partial x}\right)^2 \le 2\frac{\partial}{\partial x}\left(Au\frac{\partial u}{\partial x}\right)-\frac{\partial B}{\partial x}u^2+\frac{\partial}{\partial x}\left(Bu^2\right).$$

Now, integrating with respect to $x$ on the interval $[0,1]$, and using (7), we get:

$$\begin{aligned}\frac{d}{dt}\| u(.,t)\|^2 &\le N\| u(.,t)\|^2+B(1,t)u^2(1,t)-B(0,t)u^2(0,t)\\ &\le N\| u(.,t)\|^2+N_1 u^2(1,t)+N_2 u^2(0,t),\end{aligned}$$

where

$$\|.\|=\|.\|_{L_2([0,1])},\ N=\max_{0\le x\le 1,\,0\le t\le T}\left|\frac{\partial B(x,t)}{\partial x}\right|,\ N_1=\max_{0\le t\le T}\{0,B(1,t)\},\text{and } N_2=-\min_{0\le t\le T}\{0,B(0,t)\}.$$

We will now apply Gronwall inequality for $t\in[0,T]$ to obtain:

$$\| u(.,t)\|^2\le e^{(N+1)}\left(\|f\|^2+\int_0^t\left(N_1 u^2(1,\tau)+N_2 u^2(0,\tau)\right)d\tau\right). \tag{9}$$

It will now follow that the solution is stable with respect to $f(x)$. Furthermore, if condition (8.a) holds and $f(x)$ is identically zero, then the solution of (4) is zero, hence $f(x)$ determines the solution uniquely. Similarly if (8. b) holds, then $f(x)$ and $B(1,t)$ will uniquely determine the solution. The results for other cases are similar.

Therefore, if a solution exists, then it's unique for the problem (4)-(7) together with

(i) condition (8.a), and no other condition on $u$ is needed.
(ii) condition (8.b), and a boundary condition is needed at $x=1$.
(iii) condition (8.c), and a boundary condition is needed at $x=0$.
(iv) condition (8.d), and boundary conditions are needed at $x=0$ and $x=1$.

## THE NUMERICAL SCHEME AND ITS STABILITY:

For the above problem, several numerical schemes with relatively high rates of convergence have been constructed, and the stability of numerical solutions have been established ([8], [10]). In this work, similar numerical scheme will be constructed and the convergence of the numerical solutions to a function $u$ in the space $W_2^1\left((0,T);W_2^1(0,1)\right)$ will be proven. It will then be shown that $u$ is the weak solution of the degenerate parabolic equation. With uniqueness already known, the existence question is therefore tackled via a constructive method: As a by-product of the numerical scheme. Furthermore, some regularity results will also be obtained in the process.

We will mainly focus on problem (4)-(7), (8.b). Problems (4)-(7), (8.a) and (4)-(7), (8.d) require less work, and the case for problem (4)-(7), (8.c) is similar to that of (4)-(7), (8.b).

Let $J$ and $N$ be positive integers. Let $\Delta x = \dfrac{1}{J}$, $\Delta t = \dfrac{T}{N}$, $x_j = j\,\Delta x$, $t^n = n\,\Delta t$, $j = 0,1,\dots,J$, $n = 0,1,\dots,N$, and let the numerical solution at the point $(j\Delta x, n\Delta t)$ be denoted by $U_j^n$. For each $n$, let $U^n$ denote the vector $\left\{U_j^n \mid 0 \le j \le J\right\}$. For example, $U^0$ represents the initial condition. Also in the space of such grid functions, and for all $U^n$ and $V^n$, we define the inner product $\left\langle U^n, V^n\right\rangle = \sum_{j=0}^{J} U_j^n V_j^n \,\Delta x$, and let $\|.\|$ be the induced norm. Moreover, let $\Delta_+, \Delta_-$ and $\Delta_0$ respectively denote the forward, the backward, and the central space difference quotient operators; that is, $\Delta_+ U_j^n = \dfrac{U_{j+1}^n - U_j^n}{\Delta x}$,
$\Delta_- U_j^n = \dfrac{U_j^n - U_{j-1}^n}{\Delta x}$, and $\Delta_0 = \dfrac{\Delta_+ + \Delta_-}{2}$. Similarly, let $\Delta_t U_j^n = \dfrac{U_j^{n+1} - U_j^n}{\Delta t}$ be the time difference quotient.

To obtain second order accuracy at the boundaries, some "uniformity" will be incorporated into the scheme ([10]). This is done by first choosing a number $\lambda$ such that $0 < \lambda < \dfrac{1}{2}$, $\lambda \ge 2\,\Delta x$, and then constructing a function $\theta(x)$ on $[0,1]$ such that $\theta(x) = 1$ on $[0,\lambda]$, $\theta(x) = 0$ on $[1-\lambda, 1]$, with $\theta(x)$ being differentiable on $(0,1)$. Thus $\theta(x)$ is uniquely extended to a third degree polynomial on $[\lambda, 1-\lambda]$. Choosing $\lambda \ge 2\,\Delta x$ will ensure that $\theta(0) = \theta(\Delta x) = \theta(2\Delta x) = 1$, and $\theta(1) = \theta(1-\Delta x) = \theta(1-2\Delta x) = 0$.

For simplicity, the boundary condition $u(1,t) = \text{constant}$ will be used at $x = 1$ so that $\dfrac{d}{dt}u(1,t) = 0$. We shall now write the numerical scheme for problem (4)-(7), (8.b) :

$$\Delta_t U_j^n = \frac{1}{2}\Delta_+\left(A_{j-1/2}^{n+1/2}\;\Delta_- V_j^n\right) + \frac{1}{2}\left(1-\theta_j\right) B_j^{n+1/2}\Delta_0 V_j^n + \begin{cases} \dfrac{1}{4}\theta_j B_j^{n+1/2}\left(3\Delta_+ V_j^n - \Delta_+ V_{j+1}^n\right) & \text{if } B_j^{n+1/2} \ge 0 \\ \dfrac{1}{4}\theta_j B_j^{n+1/2}\left(3\Delta_- V_j^n - \Delta_- V_{j-1}^n\right) & \text{if } B_j^{n+1/2} < 0 \end{cases},$$

$j = 1,2,\dots,J-1$ ; $n = 0,1,\dots,N-1$,

(10)

$$V_j^n = U_j^{n+1} + U_j^n, \quad n,\ n = 0,1,\ldots,N-1,$$

$$\Delta_t U_0^n = \frac{1}{2} a_0^{n+1/2} \Delta_+ V_0^n + \frac{1}{4} B_0^{n+1/2} \left(3\Delta_+ V_0^n - \Delta_+ V_1^n\right), \quad n = 0,1,\ldots,N-1,$$

$$a(x,t) = \frac{d}{dx} A(x,t), \tag{11}$$

$$\Delta_t U_J^n = 0, \quad n = 0,1,\ldots,N-1. \tag{12}$$

Scheme (10)-(12) is second order accurate, and is stable if $B(0,t) \ge \frac{1}{7}\frac{d}{dx}A(0,t)$ ([14]). Therefore, there exists a constant $C_1$ such that for any $n,\ n = 0,1,\ldots,N-1$

$$\left\| U^n \right\| \le C_1 \tag{13}$$

(Note: A similar scheme for problem (4)-(7), (8.c) is stable if $B(1,t) \le \frac{1}{7}\frac{d}{dx}A(1,t)$ .)

## THE STABILITY OF SPACE DIFFERENCE QUOTIENTS:

For each $n,\ n = 0,1,\ldots,N-1$, let $W_j^n = U_j^{n+1} - U_j^n,\ j = 0,1,\ldots,J$ . Then the numerical scheme (10)-(12) can be written as

$$W_j^n = \frac{\Delta t}{2}\Delta_+\left(A_{j-1/2}^{n+1/2}\ \Delta_- V_j^n\right) + \frac{\Delta t}{4} B_j^{n+1/2} + \begin{cases} \left(1+2\theta_j\right)\Delta_+ V_j^n + \left(1-\theta_j\right)\Delta_+ V_{j-1}^n - \theta_j \Delta_+ V_{j+1}^n & \text{if } B_j^{n+1/2} \ge 0 \\ \left(1+2\theta_j\right)\Delta_- V_j^n + \left(1-\theta_j\right)\Delta_- V_{j+1}^n - \theta_j \Delta_- V_{j-1}^n & \text{if } B_j^{n+1/2} < 0 \end{cases}$$

$j = 1,2,\ldots,J-1$

$n = 0,1,\ldots,N-1$

(14)

$$W_0^n = \frac{\Delta t}{2} a_0^{n+1/2} \Delta_+ V_0^n + \frac{\Delta t}{4} B_0^{n+1/2}\left(3\Delta_+ V_0^n - \Delta_+ V_1^n\right), \quad n = 0,1,\ldots,N-1 \tag{15}$$

$$W_J^n = 0 \tag{16}$$

Without loss of generality, we may assume that there are indices *i* and *k*, $i < k$ , such that $B_j^{n+1/2} \ge 0$ for all $j \le i-1$ or $j \ge k$ , and $B_j^{n+1/2} < 0$ for all $i \le j \le k-1$. Next, by definition,

$$\left\|\Delta_+ U^{n+1}\right\|^2 - \left\|\Delta_+ U^n\right\|^2 = \sum_{j=0}^{J-1}\left(\frac{U_{j+1}^{n+1}-U_j^{n+1}}{\Delta x}\right)^2 \Delta x - \sum_{j=0}^{J-1}\left(\frac{U_{j+1}^{n}-U_j^{n}}{\Delta x}\right)^2 \Delta x$$

$$= \sum_{j=0}^{J-1}\frac{1}{(\Delta x)^2}\left(W_{j+1}^n - W_j^n\right)\left(V_{j+1}^n - V_j^n\right)\Delta x$$

$$= -\sum_{j=1}^{J-1}\left(\frac{V_{j+1}^n - 2V_j^n + V_{j-1}^n}{(\Delta x)^2}\right)W_j^n\,\Delta x + W_J^n\left(\frac{V_J^n - V_{J-1}^n}{\Delta x}\right) - W_0^n\left(\frac{V_1^n - V_0^n}{\Delta x}\right)$$

$$= -\sum_{j=1}^{J-1}\left(\Delta_+\Delta_- V_j^n\right)W_j^n\,\Delta x + W_J^n\Delta_- V_J^n - W_0^n\Delta_+ V_0^n.$$

Thus multiplying (14) by $-\left(\Delta_+\Delta_- V_j^n\right)\Delta x$ and summing over $j$, $j = 1, 2, ..., J-1$, multiplying (15) by $-\left(\Delta_+ V_0^n\right)$, multiplying (16) by $\left(\Delta_- V_J^n\right)$, and adding the three results, we obtain:

$$\left\|\Delta_+ U^{n+1}\right\|^2 - \left\|\Delta_+ U^n\right\|^2 = I + II + III + IV + V\,, \tag{17}$$

where

$$I = -\frac{\Delta t}{2}\sum_{j=1}^{J-1}\left(\Delta_+\Delta_- V_j^n\right)\left(\Delta_+\left(A_{j-1/2}^{n+1/2}\ \Delta_- V_j^n\right)\right)\Delta x\,, \tag{18}$$

$$II = \frac{\Delta t}{4}\sum_{j=1}^{i-1} B_j^{n+1/2}\left(-\Delta_+\Delta_- V_j^n\right)\left(\left(1+2\theta_j\right)\Delta_+ V_j^n + \left(1-\theta_j\right)\Delta_+ V_{j-1}^n - \theta_j\Delta_+ V_{j+1}^n\right)\Delta x\,, \tag{19}$$

$$III = \frac{\Delta t}{4}\sum_{j=i}^{k-1} B_j^{n+1/2}\left(-\Delta_+\Delta_- V_j^n\right)\left(\left(1+2\theta_j\right)\Delta_- V_j^n + \left(1-\theta_j\right)\Delta_- V_{j+1}^n - \theta_j\Delta_- V_{j-1}^n\right)\Delta x\,, \tag{20}$$

$$IV = \frac{\Delta t}{4}\sum_{j=K}^{J-1} B_j^{n+1/2}\left(-\Delta_+\Delta_- V_j^n\right)\left(\left(1+2\theta_j\right)\Delta_+ V_j^n + \left(1-\theta_j\right)\Delta_+ V_{j-1}^n - \theta_j\Delta_+ V_{j+1}^n\right)\Delta x\,, \tag{21}$$

$$V = \left(-\Delta_+ V_0^n\right)\left(\frac{\Delta t}{2}a_0^{n+1/2}\Delta_+ V_0^n + \frac{\Delta t}{4}B_0^{n+1/2}\left(3\Delta_+ V_0^n - \Delta_+ V_1^n\right)\right). \tag{22}$$

Next, we shall obtain upper bounds for ( $I$ )-( $V$ ):

$$I = -\frac{\Delta t}{2}\sum_{j=1}^{J-1}\left(\Delta_+\Delta_- V_j^n\right)\left(\frac{A_{j+1/2}^{n+1/2}\Delta_+ V_j^n - A_{j-1/2}^{n+1/2}\Delta_- V_j^n}{\Delta x}\right)\Delta x$$

$$=-\frac{\Delta t}{2}\sum_{j=1}^{J-1}A_j^{n+1/2}\left(\Delta_+\Delta_-V_j^n\right)^2\Delta x+\frac{\Delta t}{2}\sum_{j=1}^{J-1}\left(\Delta_+\Delta_-V_j^n\right)\left(A_j^{n+1/2}\frac{\Delta_+V_j^n-\Delta_-V_j^n}{\Delta x}-\frac{A_{j+1/2}^{n+1/2}\Delta_+V_j^n-A_{j-1/2}^{n+1/2}\Delta_-V_j^n}{\Delta x}\right)\Delta x$$

$$\le\frac{\Delta t}{2}\sum_{j=1}^{J-1}\left(\frac{\left(A_j^{n+1/2}-A_{j+1/2}^{n+1/2}\right)\left(\Delta_+V_j^n\right)^2}{\Delta x}\right)-\frac{\Delta t}{2}\sum_{j=1}^{J-1}\left(\frac{\left(A_{j-1/2}^{n+1/2}-A_j^{n+1/2}\right)\left(\Delta_-V_j^n\right)^2}{\Delta x}\right)$$

$$+\frac{\Delta t}{8}\sum_{j=1}^{J-1}\left(\frac{\left(A_{j-1/2}^{n+1/2}-2A_j^{n+1/2}+A_{j+1/2}^{n+1/2}\right)}{\left(\frac{\Delta x}{2}\right)^2}\left(\Delta_+V_j^n\right)\left(\Delta_-V_j^n\right)\right)\Delta x$$

$$=\frac{\Delta t}{8}\sum_{j=2}^{J-1}\left(\frac{\left(A_j^{n+1/2}-2A_{j+1/2}^{n+1/2}+A_{j-1}^{n+1/2}\right)\left(\Delta_-V_j^n\right)^2}{\left(\frac{\Delta x}{2}\right)^2}\right)\Delta x+\frac{\Delta t}{2}\left(\frac{\left(A_{J-1}^{n+1/2}-A_{J-1/2}^{n+1/2}\right)\left(\Delta_-V_J^n\right)^2}{\Delta x}\right)$$

$$-\frac{\Delta t}{2}\left(\frac{\left(A_{1/2}^{n+1/2}-A_1^{n+1/2}\right)\left(\Delta_-V_1^n\right)^2}{\Delta x}\right)+\frac{\Delta t}{8}\sum_{j=1}^{J-1}\left(\frac{\left(A_{j-1/2}^{n+1/2}-2A_j^{n+1/2}+A_{j+1/2}^{n+1/2}\right)}{\left(\frac{\Delta x}{2}\right)^2}\left(\Delta_+V_j^n\right)\left(\Delta_-V_j^n\right)\right)\Delta x.$$

Assuming that $a(x,t)=\frac{d}{dx}A(x,t)$ is Lipschitz continuous with respect to $x$ with Lipschitz constant $C_2$, expanding $A_{1/2}$ and $A_1$ about $x=0$, expanding $A_{J-1/2}$ and $A_{J-1}$ about $x=1$, using the inequality $\left|\Delta_+V_j^n\right|\left|\Delta_-V_j^n\right|\le\frac{1}{2}\left(\left|\Delta_+V_j^n\right|^2+\left|\Delta_-V_j^n\right|^2\right)$, and noting that $\left\|\Delta_+V^n\right\|=\left\|\Delta_-V^n\right\|$ and $\Delta_-V_1^n=\Delta_+V_0^n$, we obtain:

$$I\le\frac{7C_2\Delta t}{16}\left\|\Delta_+V^n\right\|^2+\frac{\Delta t}{4}a_0^{n+1/2}\left(\Delta_+V_0^n\right)^2-\frac{\Delta t}{4}a_J^{n+1/2}\left(\Delta_-V_J^n\right)^2\tag{23}$$

To find upper bounds for *II*, *III,* and *IV,* we do a temporary change of notation: For each *n,* we denote $\Delta_+V_j^n$ by $Y_j$. Then (19) will be written as:

$$II=\frac{\Delta t}{4}\sum_{j=1}^{i-1}B_j^{n+1/2}\left(Y_j-Y_{j-1}\right)\left(-\left(1+2\theta_j\right)Y_j-\left(1-\theta_j\right)Y_{j-1}+\theta_jY_{j+1}\right)$$
$$=\frac{\Delta t}{4}\sum_{j=1}^{i-1}B_j^{n+1/2}Q_j,$$

where

$$Q_j = \left(1-\theta_j\right)Y_{j-1}^2 - \left(1+2\theta_j\right)Y_j^2 + 3\theta_j Y_{j-1}Y_j - \theta_j Y_{j-1}Y_j + \theta_j Y_j Y_{j+1}.$$

The matrix of the quadratic form is:

$$\mathbf{Q_1} = \begin{bmatrix} 1-\theta_j & \frac{3}{2}\theta_j & \frac{-1}{2}\theta_j \\ \frac{3}{2}\theta_j & -\left(1+2\theta_j\right) & \frac{1}{2}\theta_j \\ \frac{-1}{2}\theta_j & \frac{1}{2}\theta_j & 0 \end{bmatrix},$$

which can be written as the sum of a negative definite matrix and a pseudo-null matrix $\mathbf{Q_2}$ as follows:

$$\mathbf{Q_1} = -\frac{1}{2}\theta_j \begin{bmatrix} 1 \\ -2 \\ 1 \end{bmatrix} \begin{bmatrix} 1 & -2 & 1 \end{bmatrix} + \begin{bmatrix} 1-\frac{1}{2}\theta_j & \frac{1}{2}\theta_j & 0 \\ \frac{1}{2}\theta_j & -1 & \frac{-1}{2}\theta_j \\ 0 & \frac{-1}{2}\theta_j & \frac{1}{2}\theta_j \end{bmatrix}.$$

Hence for any $\mathbf{Y} = \begin{bmatrix} Y_{j-1} & Y_j & Y_{j+1} \end{bmatrix}^T$, $\mathbf{Y^T Q_1 Y} \le \mathbf{Y^T Q_2 Y}$. Therefore since $B_j \ge 0$ in *II*, we obtain:

$$II \le \frac{\Delta t}{4}\sum_{j=1}^{i-1} B_j^{n+1/2}\left(\left(1-\frac{1}{2}\theta_j\right)Y_{j-1}^2 - Y_j^2 + \frac{1}{2}\theta_j Y_{j+1}^2 + \theta_j Y_{j-1}Y_j - \theta_j Y_j Y_{j+1}\right)$$

$$= \frac{\Delta t}{4}\left(\sum_{j=0}^{i-2} B_{j+1}^{n+1/2}\left(1-\frac{1}{2}\theta_{j+1}\right)Y_j^2 - \sum_{j=1}^{i-1} B_j^{n+1/2}Y_j^2 + \sum_{j=2}^{i}\frac{1}{2}B_{j-1}^{n+1/2}\theta_{j-1}Y_j^2 + \sum_{j=1}^{i-1} B_j^{n+1/2}\theta_j Y_j\left(Y_{j-1}-Y_{j+1}\right)\right).$$

If the same procedure is repeated for *IV* and a similar procedure for *III*, and all results are put together in (17), inequality (24) will be resulted. In eliminating some of the boundary terms, we had to make the assumption that: $\frac{d}{dx}A(1,t) + B(1,t) \ge 0$, which is not very severe and, in fact, from the applications point of view, we have: $\frac{d}{dx}A(1,t) = 0$.

$$\left\|\Delta_+U^{n+1}\right\|^2-\left\|\Delta_+U^n\right\|^2=C_5\Delta t\left\|\Delta_+V^n\right\|^2-\frac{\Delta t}{4}B_0^{n+1/2}\left(2Y_0^2-2Y_0Y_1+\frac{1}{2}Y_1^2\right)$$
$$=C_5\Delta t\left\|\Delta_+V^n\right\|^2-\frac{\Delta t}{4}B_0^{n+1/2}\begin{bmatrix}Y_0 & Y_1\end{bmatrix}\begin{bmatrix}\sqrt{2}\\ -\frac{\sqrt{2}}{2}\end{bmatrix}\begin{bmatrix}\sqrt{2} & -\frac{\sqrt{2}}{2}\end{bmatrix}\begin{bmatrix}Y_0\\ Y_1\end{bmatrix}, \tag{24}$$

where $C_5=C_5(C_2,C_3,C_4)$, and $C_3$ and $C_4$ are Lipschitz constants (with respect to *x) of* functions $B(x,t)$ and $\theta(x)B(x,t)$ respectively. Therefore,

$$\left\|\Delta_+U^{n+1}\right\|^2-\left\|\Delta_+U^n\right\|^2\le C_5\Delta t\left\|\Delta_+V^n\right\|^2, \tag{25}$$

or,

$$\left\|\Delta_+U^{n+1}\right\|^2\le\frac{1+C_5\Delta t}{1-C_5\Delta t}\left\|\Delta_+U^n\right\|^2 \tag{26}$$

It follows from (26) that if $\frac{d}{dx}u(x,0)\in L_2\left(\left[0,1\right]\right)$ and $\Delta t<\frac{1}{C_5}$, then there is a constant $C_6$ such that for all $n$, $n=0,1,...,N-1$,

$$\left\|\Delta_+U^n\right\|\le C_6 \tag{27}$$

( Note: A similar appropriate scheme for problem (4)-(7), (8.c) would have stable space difference quotients if $\frac{d}{dx}A(0,t)+B(0,t)\le 0$ and $\Delta t$ is chosen appropriately.)

The following is a review of the list of constants already used, and a few new constants that will be used in the sequel:

$C_1=$bound for $\left\|U^n\right\|$

$C_2=$ Lipschitz constant of $\frac{d}{dx}A(x,t)$ with respect to $x$

$C_3=$ Lipschitz constant of $B(x,t)$ with respect to $x$

$C_4=$ Lipschitz constant of $\theta(x)B(x,t)$ with respect to $x$

$C_5=C_5(C_2,C_3,C_4)$

$C_6=$bound for $\left\|\Delta_+U^n\right\|$ (or, $\left\|\Delta_-U^n\right\|$)

$C_7=$ Lipschitz constant of $A(x,t)$ with respect to t (or, the bound for $\left|\frac{d}{dt}A(x,t)\right|$)

$C_8 =$ Lipschitz constant of $A(x,t)$ with respect to $x$ (or, the bound for $\left|\frac{d}{dx}A(x,t)\right|$)

$C_9 = \max_{0\le x\le 1,\, 0\le t\le T}\left\{\left|A(x,t)\right|\right\}$

$C_{10} = \max_{0\le x\le 1,\, 0\le t\le T}\left\{\left|B(x,t)\right|\right\}$

$C_{11} =$ bound for $\sum_{n=0}^{N-1}\left\|\Delta_t U^n\right\|^2 \Delta t$

**THE STABILITY OF TIME DIFFERENCE QUOTIENTS:**

Using the numerical scheme (10)-(12), we multiply (10) by $\Delta_t U_j^n \Delta x\,\Delta t$ and sum over $j$ and $n$, $j = 1,2,...,J-1$, $n = 0,1,...,N-1$; and multiply (11) and (12) respectively by $\Delta_t U_0^n \Delta x\,\Delta t$ and $\Delta_t U_J^n \Delta x\,\Delta t$ and sum over $n$, $n = 0,1,...,N-1$, to obtain:

$$\sum_{n=0}^{N-1}\left\|\Delta_t U^n\right\|^2 \Delta t = VI + VII + VIII + IX + X\,, \tag{28}$$

where,

$$VI = \sum_{n=0}^{N-1}\sum_{j=1}^{J-1}\frac{1}{2}\left(\Delta_t U_j^n\right)\left(\Delta_+\left(A_{j-1/2}^{n+1/2}\ \Delta_- V_j^n\right)\right)\Delta x\,\Delta t, \tag{29}$$

$$VII = \sum_{n=0}^{N-1}\sum_{j=1}^{i-1}\frac{1}{4}B_j^{n+1/2}\left(\Delta_t U_j^n\right)\left(\left(1+2\theta_j\right)\Delta_+ V_j^n + \left(1-\theta_j\right)\Delta_+ V_{j-1}^n - \theta_j\Delta_+ V_{j+1}^n\right)\Delta x\,\Delta t, \tag{30}$$

$$VIII = \sum_{n=0}^{N-1}\sum_{j=i}^{k-1}\frac{1}{4}B_j^{n+1/2}\left(\Delta_t U_j^n\right)\left(\left(1+2\theta_j\right)\Delta_- V_j^n + \left(1-\theta_j\right)\Delta_- V_{j+1}^n - \theta_j\Delta_- V_{j-1}^n\right)\Delta x\,\Delta t, \tag{31}$$

$$IX = \sum_{n=0}^{N-1}\sum_{j=k}^{J-1}\frac{1}{4}B_j^{n+1/2}\left(\Delta_t U_j^n\right)\left(\left(1+2\theta_j\right)\Delta_+ V_j^n + \left(1-\theta_j\right)\Delta_+ V_{j-1}^n - \theta_j\Delta_+ V_{j+1}^n\right)\Delta x\,\Delta t, \tag{32}$$

$$X = \sum_{n=0}^{N-1}\left(\Delta_t U_0^n\right)\left(\frac{1}{2}a_0^{n+1/2}\Delta_+ V_0^n + \frac{1}{4}B_0^{n+1/2}\left(3\Delta_+ V_0^n - \Delta_+ V_1^n\right)\right)\Delta x\,\Delta t\,. \tag{33}$$

Similar algebraic manipulations as those in the previous section and an $\varepsilon$-inequality type argument yields:

$$VI \le \frac{1}{2}C_7 C_6^2 T - \frac{1}{2}\sum_{j=0}^{J-1}\left(\sum_{N=0}^{N-1}A_{j+1/2}^{n+1}\left(\Delta_+ U_j^{n+1}\right)^2 - \sum_{N=0}^{N-1}A_{j+1/2}^{n}\left(\Delta_+ U_j^{n}\right)^2\right)\Delta x$$
$$+\frac{1}{8}C_8\sum_{n=0}^{N-1}\left(\varepsilon\left(\Delta_+ V_0^n\right)^2 + \frac{1}{\varepsilon}\left(\Delta_t U_0^n\right)^2 + \varepsilon\left(\Delta_- V_J^n\right)^2 + \frac{1}{\varepsilon}\left(\Delta_t U_J^n\right)^2\right)\Delta x\,\Delta t$$

$$\leq \frac{1}{2}C_7 C_6^2 T + \frac{1}{2}C_9 \left\|U^0\right\|^2 + \frac{1}{8}C_8 \varepsilon T\left(\left(\Delta_- V_1^n\right)^2 + \left(\Delta_- V_J^n\right)^2\right)\Delta x + \frac{1}{8\varepsilon}C_8 \sum_{n=0}^{N-1}\left(\left(\Delta_t U_0^n\right)^2 + \left(\Delta_t U_J^n\right)^2\right)\Delta x\,\Delta t\,.$$

Considering the fact that $|\theta(x)| \leq 1$ for all $x \in [0,1]$, other terms in (28), are put together as shown below:

$$VII + VIII + IX + X \leq \frac{1}{4}C_{10}\sum_{n=0}^{N-1}\sum_{j=1}^{i-1}\left(3\varepsilon\left(\Delta_+ V_j^n\right)^2 + \varepsilon\left(\Delta_+ V_{j+1}^n\right)^2 + \varepsilon\left(\Delta_+ V_{j-1}^n\right)^2 + \frac{5}{\varepsilon}\left(\Delta_t U_j^n\right)^2\right)\Delta x\,\Delta t$$

$$+\frac{1}{4}C_{10}\sum_{n=0}^{N-1}\sum_{j=i}^{k-1}\left(3\varepsilon\left(\Delta_- V_j^n\right)^2 + \varepsilon\left(\Delta_- V_{j+1}^n\right)^2 + \varepsilon\left(\Delta_- V_{j-1}^n\right)^2 + \frac{5}{\varepsilon}\left(\Delta_t U_j^n\right)^2\right)\Delta x\,\Delta t$$

$$+\frac{1}{4}C_{10}\sum_{n=0}^{N-1}\sum_{j=k}^{J-1}\left(3\varepsilon\left(\Delta_+ V_j^n\right)^2 + \varepsilon\left(\Delta_+ V_{j+1}^n\right)^2 + \varepsilon\left(\Delta_+ V_{j-1}^n\right)^2 + \frac{5}{\varepsilon}\left(\Delta_t U_j^n\right)^2\right)\Delta x\,\Delta t$$

$$+\frac{1}{4}C_{10}\sum_{n=0}^{N-1}\left(3\varepsilon\left(\Delta_+ V_0^n\right)^2 + \varepsilon\left(\Delta_+ V_1^n\right)^2 + \frac{4}{\varepsilon}\left(\Delta_t U_J^n\right)^2\right)\Delta x\,\Delta t$$

$$+\frac{1}{2}C_8\sum_{n=0}^{N-1}\left(\varepsilon\left(\Delta_+ V_0^n\right)^2 + \frac{1}{\varepsilon}\left(\Delta_t U_J^n\right)^2\right)\Delta x\,\Delta t.$$

They are then combined with *VI* to get:

$$\sum_{n=0}^{N-1}\left\|\Delta_t U^n\right\|^2 \Delta t \leq \frac{1}{2}C_7 C_6^2 T + \frac{1}{2}C_9 C_6{}^2 + \frac{3}{8}C_8 \varepsilon T C_6{}^2 + \frac{1}{8\varepsilon}\left(3C_8 + 10C_{10}\right)\sum_{n=0}^{N-1}\left\|\Delta_t U^n\right\|^2 \Delta t$$

We may now choose $\varepsilon = \left(3C_8 + 10C_{10}\right)$, and arrive at the desired conclusion:

$$\sum_{n=0}^{N-1}\left\|\Delta_t U^n\right\|^2 \Delta t \leq C_{11}\,, \tag{34}$$

$$\left(C_{11} = \frac{4}{7}C_7 C_6^2 T + \frac{4}{7}C_9 C_6{}^2 + \frac{3}{7}C_8 T C_6{}^2\left(3C_8 + 10C_{10}\right)\right).$$

With inequalities (13), (27), and (34) at our disposal, we are now in the position to tackle the question of existence via a constructive method.

## QUALITATIVE PROPERTIES OF THE SOLUTION:

In the sequel, we will use the following inner product and the corresponding energy norm: For all integrable functions $u(x,t)$ and $v(x,t)$ on $[0,1]\times[0,T]$, we define:

$$\langle u, v\rangle_E = \int_0^T \int_0^1 u\, v\, dx\, dt, \text{ and } \|u\|_E^2 = \int_0^T \|u\|_{L_2([0,1])}^2 dt.$$

Also for any set $S$, let $\chi_S$ denote its characteristic function; that is,

$$\chi_S(\omega) = \begin{cases} 1 & \text{if } \omega \in S \\ 0 & \text{if } \omega \notin S \end{cases}$$

We will now consider the degenerate parabolic equation (1) along with an arbitrary initial condition, such as $u(x,0) = f(x)$, and solve it numerically by the numerical scheme (10)-(12). Note that since $B(0,t) = 0$ and $B(1,t) = 0$, then this equation falls into the type with boundary characteristic (8.a). Therefore scheme (10)-(12) has to be slightly adjusted accordingly. To obtain a numerical solution, we assigned the following values to the parameters in the equation: $\sigma = 0.05,\ r = 0.05, D_0 = 0.1,\ T = 1$ unit.. For the initial condition, $f(x) = \tanh(x)$ was chosen. We will next use the numerical solution $U_j^n$ to construct Step functions $\bar{r}(x,t)$ and $\bar{l}(x,t)$ as follows:

$$\begin{aligned}
&\bar{r}(x,t) = \sum_{n=0}^{N-1}\sum_{j=1}^{J} \chi_{(D]_j} U_j^{n+1}, && (x,t) \in (0,1]\times(0,T],\\
&\bar{r}(0,t) = U_0^{n+1}, && t \in (n\Delta t, (n+1)\Delta t],\ n = 0,1,\dots,N-1,\\
&\bar{r}(x,0) = U_j^0, && x \in ((j-1)\Delta x, j\Delta x],\ j = 1,2,\dots,J,\\
&\bar{r}(0,0) = U_0^0,\\
&(D]_j = \{(x,t) : x \in ((j-1)\Delta x, j\Delta x], t \in (n\Delta t, (n+1)\Delta t], j = 1,2,\dots,J, n = 0,1,\dots,N-1\}.\\
&\bar{l}(x,t) = \sum_{n=0}^{N-1}\sum_{j=0}^{J-1} \chi_{[D)_j} U_j^{n+1}, && (x,t) \in [0,1)\times(0,T],\\
&\bar{l}(1,t) = U_J^{n+1}, && t \in (n\Delta t, (n+1)\Delta t],\ n = 0,1,\dots,N-1,\\
&\bar{l}(x,0) = U_j^0, && x \in [j\Delta x, (j+1)\Delta x),\ j = 0,1,\dots,J-1,\\
&\bar{l}(1,0) = U_J^0,\\
&[D)_j = \{(x,t) : x \in [j\Delta x, (j+1)\Delta x); t \in (n\Delta t, (n+1)\Delta t]; j = 0,1,\dots,J-1; n = 0,1,\dots,N-1\}.
\end{aligned}$$

The graph of $\bar{l}(x,t)$ is shown below. The graph of $\bar{r}(x,t)$ is similar with the difference that, for each rectangular sub-domain $\Delta x \times \Delta t$, $\bar{r}(x,t)$ attains the solution values at the larger space corners.

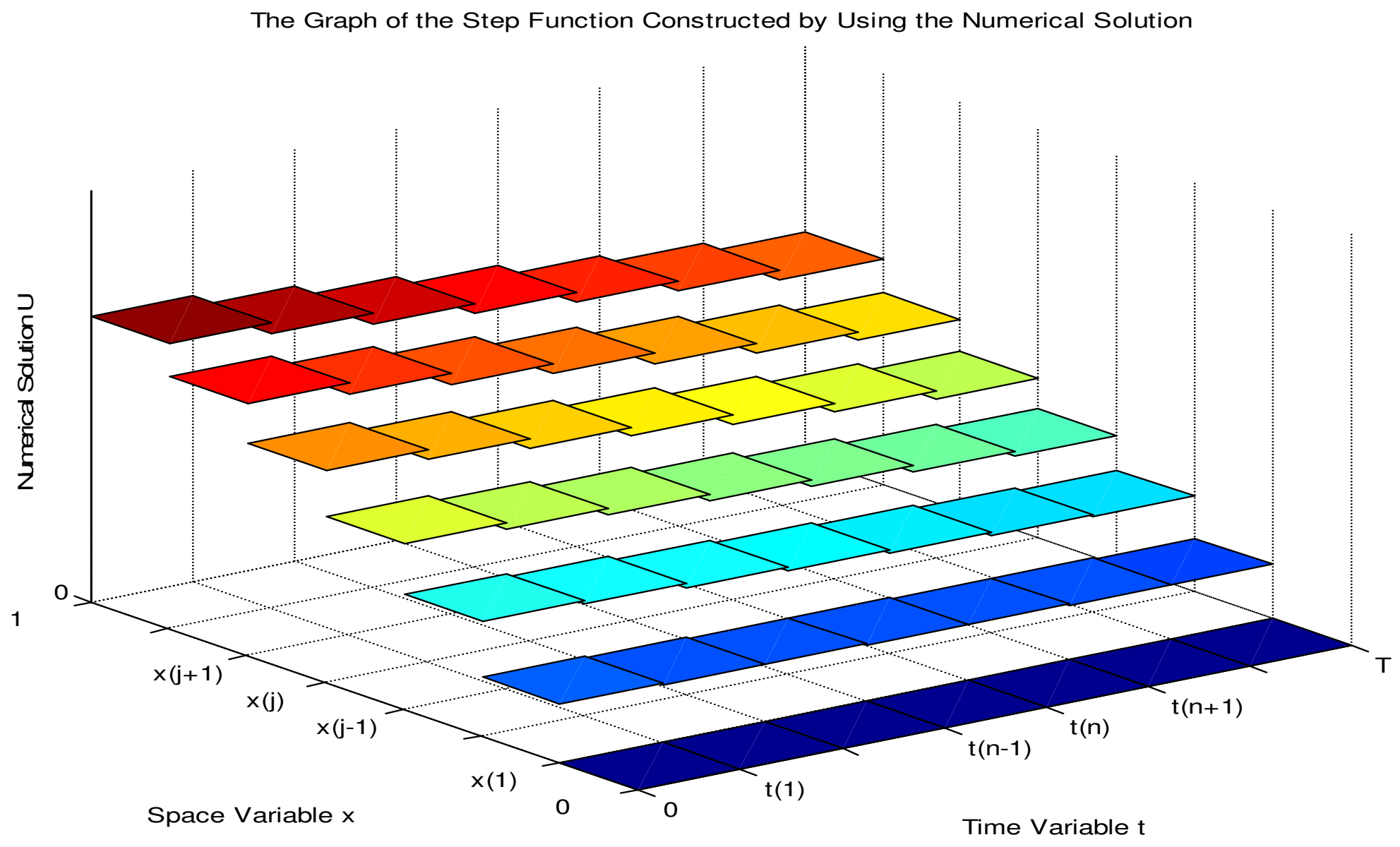

We will now derive some qualitative properties of $\bar{r}(x,t)$ and $\bar{l}(x,t)$: For any fixed *n*,

$$\left\|\bar{r}(x,t)\right\|^2_{L_2([0,1])} = \int_0^1 \sum_{j=1}^{J} \chi_{(D]_j} \left(U_j^{n+1}\right)^2 dx = \sum_{j=0}^{J} \left(U_j^{n+1}\right)^2 \int_{x\in(D]_j} dx = \left\|U^{n+1}\right\|^2 \le \left(C_1\right)^2.$$

Similarly, $\left\|\bar{l}(x,t)\right\|^2_{L_2([0,1])} \le \left(C_1\right)^2$. Therefore, for any $t \in [0,T]$,

$\bar{r}(x,.)$ , $\bar{l}(x,.) \in L_2\left([0,1]\right)$.

Moreover, since the above is true for all *t*, it results that $\sup_t \left\|\bar{r}(x,t)\right\|_{L_2([0,1])} \le C_1$, and $\sup_t \left\|\bar{l}(x,t)\right\|_{L_2([0,1])} \le C_1$. That is, $\bar{r}$ and $\bar{l} \in L_\infty\left((0,T); L_2\left([0,1]\right)\right)$. In fact, from $\left\|\bar{r}(x,t)\right\|^2_{L_2\left([0,T];L_2([0,1])\right)} = \left\|\bar{r}(x,t)\right\|^2_E = \int_0^T \left\|\bar{r}(x,t)\right\|^2_{L_2([0,1])} dt \le \left(C_1\right)^2 T$, it follows that:

$\bar{r}(x,t)$ , $\bar{l}(x,t) \in L_2\left([0,T]; L_2\left([0,1]\right)\right)$.

That is, there is a constant $C_1' = C_1\sqrt{T}$ such that

$$\left\|\bar{r}(x,t)\right\|_E \text{ , } \left\|\bar{l}(x,t)\right\|_E \le C_1'. \tag{35}$$

Thus in the Hilbert space $L_2\left(\left[0,T\right];L_2\left(\left[0,1\right]\right)\right)$, inequality (35) gives rise to a compactness in weak * topology. Since a Hilbert space is reflexive, a weak compactness follows immediately. To be more precise, let $\Phi = C^1\left(\left(0,T\right);C^1\left(0,1\right)\right)$ be a set of test functions. Then there exists a sequence $\left\{\Delta x_i, \Delta t_i\right\}$ such that as $\left\{\Delta x_i, \Delta t_i\right\} \to 0$, the functions $\bar{r}(x,t)$ and $\bar{l}(x,t)$ respectively converge weakly to some functions $r(x,t)$ and $l(x,t)$ in $L_2\left(\left[0,T\right];L_2\left(\left[0,1\right]\right)\right)$; that is, for all $\phi \in \Phi$,

$$\begin{aligned} \lim_{(\Delta x,\Delta t)\to 0} \left\langle \bar{r}(x,t), \phi(x,t) \right\rangle_E &= \left\langle r(x,t), \phi(x,t) \right\rangle_E , \\ \lim_{(\Delta x,\Delta t)\to 0} \left\langle \bar{l}(x,t), \phi(x,t) \right\rangle_E &= \left\langle l(x,t), \phi(x,t) \right\rangle_E . \end{aligned} \tag{36}$$

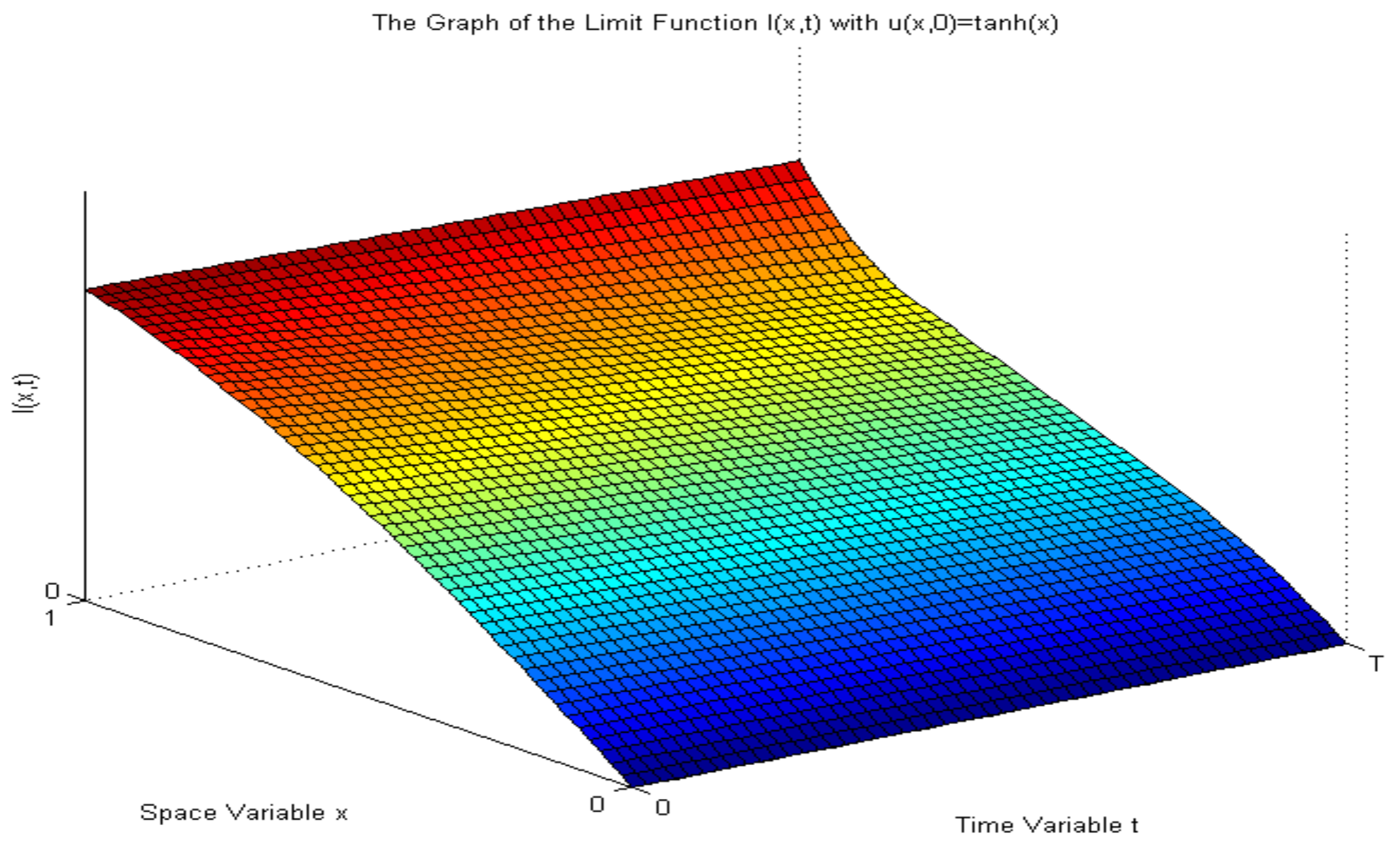

We will next utilize the space and the time difference quotients obtained from the numerical solution to construct step functions $\bar{r}'(x,t)$, $\bar{l}'(x,t)$, $\bar{r}_t(x,t)$, and $\bar{l}_t(x,t)$ as follows:

$$\bar{r}'(x,t) = \sum_{n=0}^{N-1}\sum_{j=0}^{J-1} \chi_{(D]_{j+1}}\left(\Delta_+ U_j^{n+1}\right), \qquad (x,t)\in\left(0,1\right]\times\left(0,T\right], \quad \bar{r}'(x,0) = \Delta_+ U_j^0 .$$

$$\bar{l}'(x,t) = \sum_{n=0}^{N-1}\sum_{j=1}^{J} \chi_{[D)_{j-1}}\left(\Delta_- U_j^{n+1}\right), \qquad (x,t)\in\left[0,1\right)\times\left(0,T\right], \quad \bar{l}'(x,0) = \Delta_- U_j^0 .$$

$$\bar{r}_t(x,t) = \sum_{n=0}^{N-1}\sum_{j=1}^{J} \chi_{(D]_j}\left(\Delta_t U_j^{n}\right), \qquad (x,t)\in\left(0,1\right]\times\left(0,T\right], \quad \bar{r}_t(0,t) = \Delta_t U_0^n .$$

$$\bar{l}_t(x,t)=\sum_{n=0}^{N-1}\sum_{j=0}^{J-1}\chi_{[D)_j}\left(\Delta_t U_j^n\right), \qquad (x,t)\in[0,1)\times(0,T], \quad \bar{l}_t(1,t)=\Delta_t U_J^n \ .$$

The functions constructed above, have certain qualitative properties that we will explore. These properties will play an important role in establishing the existence of the solution to problem (1), and revealing some of its regularity properties. We will first show that the above functions belong to the same function spaces as functions $\bar{r}(x,t)$ and $\bar{l}(x,t)$:

$$\begin{aligned}\left\|\bar{r}'(x,t)\right\|^2_{L_2([0,T];L_2([0,1]))} &= \left\|\bar{r}'(x,t)\right\|^2_E = \int_0^T\int_0^1\left(\sum_{n=0}^{N-1}\sum_{j=0}^{J-1}\chi_{(D]_{j+1}}\left(\Delta_+ U_j^{n+1}\right)\right)^2 dx\,dt \\ &= \sum_{n=0}^{N-1}\sum_{j=0}^{J-1}\left(\Delta_+ U_j^{n+1}\right)^2 \int\limits_{x\in(j\Delta x,(j+1)\Delta x]} dx \int\limits_{t\in(n\Delta t,(n+1)\Delta t]} dt \\ &= \sum_{n=0}^{N-1}\left(\sum_{j=0}^{J-1}\left(\Delta_+ U_j^{n+1}\right)^2\Delta x\right)\Delta t = \sum_{n=0}^{N-1}\left\|\Delta_+ U^{n+1}\right\|^2\Delta t \le \left(C_6\right)^2 T.\end{aligned}$$

Also,

$$\begin{aligned}\left\|\bar{r}_t(x,t)\right\|^2_E &= \int_0^T\int_0^1\left(\sum_{n=0}^{N-1}\sum_{j=1}^{J}\chi_{(D]_j}\left(\Delta_t U_j^n\right)\right)^2 dx\,dt \\ &= \sum_{n=0}^{N-1}\sum_{j=1}^{J}\left(\Delta_t U_j^n\right)^2 \int\limits_{x\in((j-1)\Delta x,\, j\Delta x]} dx \int\limits_{t\in(n\Delta t,(n+1)\Delta t]} dt \\ &= \sum_{n=0}^{N-1}\left(\sum_{j=1}^{J}\left(\Delta_t U_j^n\right)^2\Delta x\right)\Delta t = \sum_{n=0}^{N-1}\left\|\Delta_t U^n\right\|^2\Delta t \le C_{11}.\end{aligned}$$

Same respective results also hold for $\bar{l}'(x,t)$ and $\bar{l}_t(x,t)$, hence establishing the existence of constants $C_2'$ ($C_2' = C_6\sqrt{T}$) and $C_3'$ ($C_3' = \sqrt{C_{11}}$) such that

$$\left\|\bar{r}'(x,t)\right\|_E \ , \ \left\|\bar{l}'(x,t)\right\|_E \le C_2' \ . \tag{37}$$

$$\left\|\bar{r}_t(x,t)\right\|_E \le \ , \ \left\|\bar{l}_t(x,t)\right\|_E \le C_3' \ . \tag{38}$$

From (37) and (38), it follows that there exist sequences $\{\Delta x_i, \Delta t_i\}$ and $\{\Delta x_k, \Delta t_k\}$ such that as $\{\Delta x_i, \Delta t_i\}\to 0$ and $\{\Delta x_k, \Delta t_k\}\to 0$ respectively, the pairs $\left(\bar{r}'(x,t), \bar{l}'(x,t)\right)$ and $\left(\overline{r_t}(x,t), \overline{l_t}(x,t)\right)$ converge weakly to some pairs $\left(r'(x,t), l'(x,t)\right)$ and $\left(r_t(x,t), l_t(x,t)\right)$ in $L_2\left([0,T]; L_2\left([0,1]\right)\right)$; that is, for any $\phi \in \Phi$,

$$\lim_{(\Delta x,\Delta t)\to 0} \left\langle \overline{r}'(x,t), \phi(x,t)\right\rangle_E = \left\langle r'(x,t), \phi(x,t)\right\rangle_E ,$$
$$\lim_{(\Delta x,\Delta t)\to 0} \left\langle \overline{l}'(x,t), \phi(x,t)\right\rangle_E = \left\langle l'(x,t), \phi(x,t)\right\rangle_E , \tag{39}$$

and,

$$\lim_{(\Delta x,\Delta t)\to 0} \left\langle \overline{r_t}(x,t), \phi(x,t)\right\rangle_E = \left\langle r_t(x,t), \phi(x,t)\right\rangle_E ,$$
$$\lim_{(\Delta x,\Delta t)\to 0} \left\langle \overline{l_t}(x,t), \phi(x,t)\right\rangle_E = \left\langle l_t(x,t), \phi(x,t)\right\rangle_E . \tag{40}$$

**Lemma 1:**
Let $D_x$ and $D_t$ denote the generalized derivatives in the space $L_2\left((0,T); L_2(0,1)\right)$ with respect to *x* and *t* respectively. Then (a) $l'(x,t) = D_x\left(r(x,t)\right)$, and $r'(x,t) = D_x\left(l(x,t)\right)$. (b) $r_t(x,t) = D_t\left(r(x,t)\right)$, and $l_t(x,t) = D_t\left(l(x,t)\right)$.

**Proof (a):**
Let $\phi(x,t) \in \Phi$ be any test function with compact support on $[0,T]\times[0,1]$, and let $\phi_j^n = \phi(j\Delta x, n\Delta t)$. Then by summation by parts,

$$\begin{aligned}
\left\langle r(x,t), \frac{d}{dx}\phi(x,t)\right\rangle_E &= \lim_{(\Delta x,\Delta t)\to 0} \left\langle \overline{r}(x,t), \frac{d}{dx}\phi(x,t)\right\rangle_E \\
&= \lim_{(\Delta x,\Delta t)\to 0} \int_0^T \int_0^1 \left( \sum_{n=0}^{N-1} \sum_{j=1}^{J} \chi_{(D]_j} U_j^{n+1} \right)\left( \lim_{\Delta x\to 0} \frac{\phi_{j+1}^{n+1} - \phi_j^{n+1}}{\Delta x} \right) dx\, dt \\
&= \lim_{(\Delta x,\Delta t)\to 0} \sum_{n=0}^{N-1} \left( (-1) \sum_{j=1}^{J} \phi_j^{n+1} \left( \frac{U_j^{n+1} - U_{j-1}^{n+1}}{\Delta x} \right) \right) \Delta x\, \Delta t \\
&= -\lim_{(\Delta x,\Delta t)\to 0} \int_0^T \int_0^1 \left( \sum_{n=0}^{N-1} \sum_{j=1}^{J} \chi_{[D)_j} \left(\Delta_U_j^{n+1}\right) \right) \phi(x,t)\, dx\, dt \\
&= -\lim_{(\Delta x,\Delta t)\to 0} \int_0^T \int_0^1 \overline{l}'(x,t)\ \phi(x,t)\, dx\, dt = -\int_0^T \int_0^1 l'(x,t)\ \phi(x,t)\, dx\, dt \\
&= -\left\langle l'(x,t), \phi(x,t)\right\rangle_E .
\end{aligned}$$

Similarly, $\left\langle l(x,t), \frac{d}{dx}\phi(x,t) \right\rangle_E = -\left\langle r'(x,t), \phi(x,t) \right\rangle_E$ .

**Proof (b):**

$$\left\langle r(x,t), \frac{d}{dt}\phi(x,t) \right\rangle_E = \lim_{(\Delta x,\Delta t)\to 0} \left\langle \bar{r}(x,t), \frac{d}{dt}\phi(x,t) \right\rangle_E$$

$$= \lim_{(\Delta x,\Delta t)\to 0} \int_0^T \int_0^1 \left( \sum_{n=0}^{N-1} \sum_{j=1}^{J} \chi_{(D]_j} U_j^{n+1} \right) \left( \lim_{\Delta t\to 0} \frac{\phi_j^{n+1} - \phi_j^n}{\Delta t} \right) dx\, dt$$

$$= \lim_{(\Delta x,\Delta t)\to 0} \sum_{n=0}^{N-1} \left( (-1) \sum_{j=1}^{J} \phi_j^n \left( \frac{U_j^{n+1} - U_j^n}{\Delta x} \right) \right) \Delta x\, \Delta t$$

$$= - \lim_{(\Delta x,\Delta t)\to 0} \int_0^T \int_0^1 \left( \sum_{n=0}^{N-1} \sum_{j=0}^{J-1} \chi_{(D]_j} \left( \Delta_t U_j^n \right) \right) \left( \lim_{(\Delta x,\Delta t)\to 0} \phi_j^n \right) dx\, dt$$

$$= - \lim_{(\Delta x,\Delta t)\to 0} \int_0^T \int_0^1 \bar{r}_t(x,t)\ \phi(x,t)\, dx\, dt = -\int_0^T \int_0^1 r_t(x,t)\ \phi(x,t)\, dx\, dt$$

$$= -\left\langle r_t(x,t), \phi(x,t) \right\rangle_E .$$

Similarly, $\left\langle l(x,t), \frac{d}{dt}\phi(x,t) \right\rangle_E = -\left\langle l_t(x,t), \phi(x,t) \right\rangle_E$ .

■

In light of lemma 1 and the previous qualitative observations, it results that the functions $r(x,t)$ and $l(x,t)$ belong to the Sobolev space $W_2^1\left((0,T);W_2^1(0,1)\right)$. Naturally, we will next want to show that $r(x,t)$ and $l(x,t)$ are in fact the same. To do so, we will need the result of the following corollary:

**Corollary 1**:
For any fixed *n*, and for any grid-function $V^n = \left\{ V_j^n \mid 0 \le j \le J \right\}$ on the interval $[0,1]$,

$$\max_{0\le j\le J} \left| V_j^n \right|^2 \le \left\| V^n \right\| \left( 2 \left\| \Delta_+ V^n \right\| + \left\| V^n \right\| \right). \tag{41}$$

**Proof:**
For any fixed *n* on the interval $[0,1]$, Let *i* and *j* be two indices, each of which are not necessarily unique, such that $V_i^n = \min_j \left| V_j^n \right|$ and $V_k^n = \max_j \left| V_j^n \right|$. Then without loss of generality,

$$\sum_{j=i}^{k} V_j^n \left(\Delta_+ V_j^n\right)\Delta x = \sum_{j=i+1}^{k+1} V_j^n V_{j-1}^n - \sum_{j=i}^{k}\left(V_j^n\right)^2 = -\sum_{j=i+1}^{k} V_j^n \left(\Delta_- V_j^n\right)\Delta x + V_k^n V_{k+1}^n - \left(V_i^n\right)^2 + \left(V_k^n\right)^2 - \left(V_k^n\right)^2$$

Thus by Cauchy-Schwartz inequality,

$$\left(V_k^n\right)^2 \le \sum_{j=i}^{k-1}\left|V_j^n\right|\left|\Delta_+ V_j^n\right|\Delta x + \sum_{j=i+1}^{k}\left|V_j^n\right|\left|\Delta_- V_j^n\right|\Delta x + \left(V_i^n\right)^2$$

$$\le \sum_{j=0}^{J-1}\left|V_j^n\right|\left|\Delta_+ V_j^n\right|\Delta x + \sum_{j=1}^{J}\left|V_j^n\right|\left|\Delta_- V_j^n\right|\Delta x + \frac{\sum_{j=0}^{J}\left|V_j^n\right|^2 \Delta x}{1-0}$$

$$\le \left\|V^n\right\|\left\|\Delta_+ V^n\right\| + \left\|V^n\right\|\left\|\Delta_- V^n\right\| + \left\|V^n\right\|^2 .$$

This yields the desired result.

■

**Lemma 2:**
$r(x,t) = l(x,t)$.

**Proof:**
We will show that $r(x,t)$ and $l(x,t)$ are both equal to a third function $s(x,t)$ whose existence will follow from the following construction: In each rectangle $\left(D\right]_j^n$ of the mesh, let $P_j^n$ denote the interpolating surface which is linear in both *x* and *t* directions with values at the four nodal points equal to those of the numerical solution. The graph below is that of $P_{j+1}^n$.

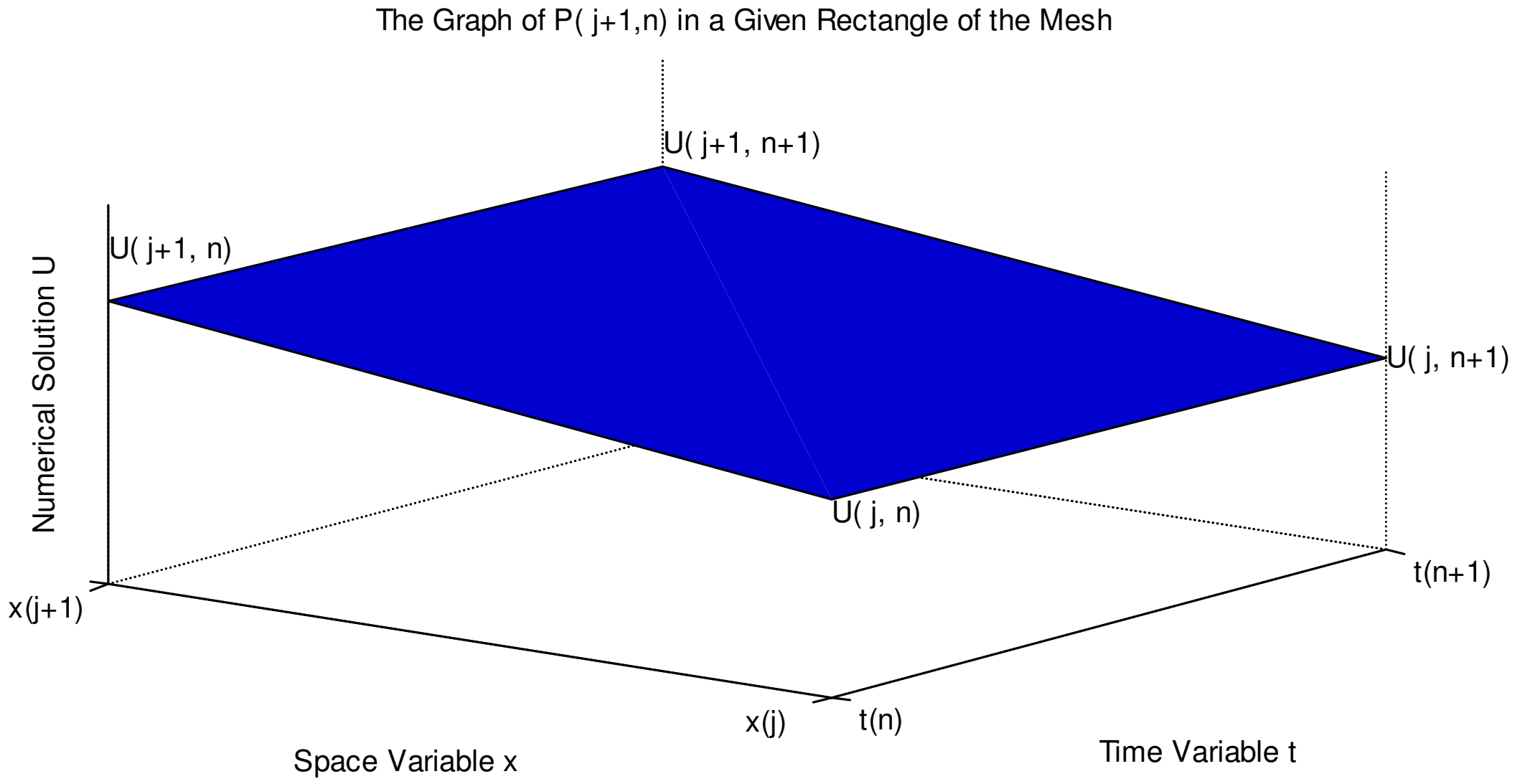

We will then construct the function $\bar{s}(x,t)$ on $[0,T]\times[0,1]$ as the collection of all $P_j^n$'s . In other words $\bar{s}(x,t)$ is the linear interpolation of the numerical solution values.

$$\bar{s}(x,t)=\sum_{n=0}^{N-1}\sum_{j=1}^{J}\chi_{(D]_j}P_j^n, \qquad j=1,2,...,J;\ n=0,1,...N-1,$$

$$\bar{s}(0,t)=U_0^n, \qquad t\in[n\Delta t,(n+1)\Delta t],\ n=0,1,...,N-1,$$

$$\bar{s}(x,0)=U_j^0, \qquad x\in[(j-1)\Delta x, j\Delta x],\ j=1,2,...,J,$$

$$\bar{s}(1,t)=U_J^n, \qquad t\in[n\Delta t,(n+1)\Delta t],\ n=0,1,...,N-1.$$

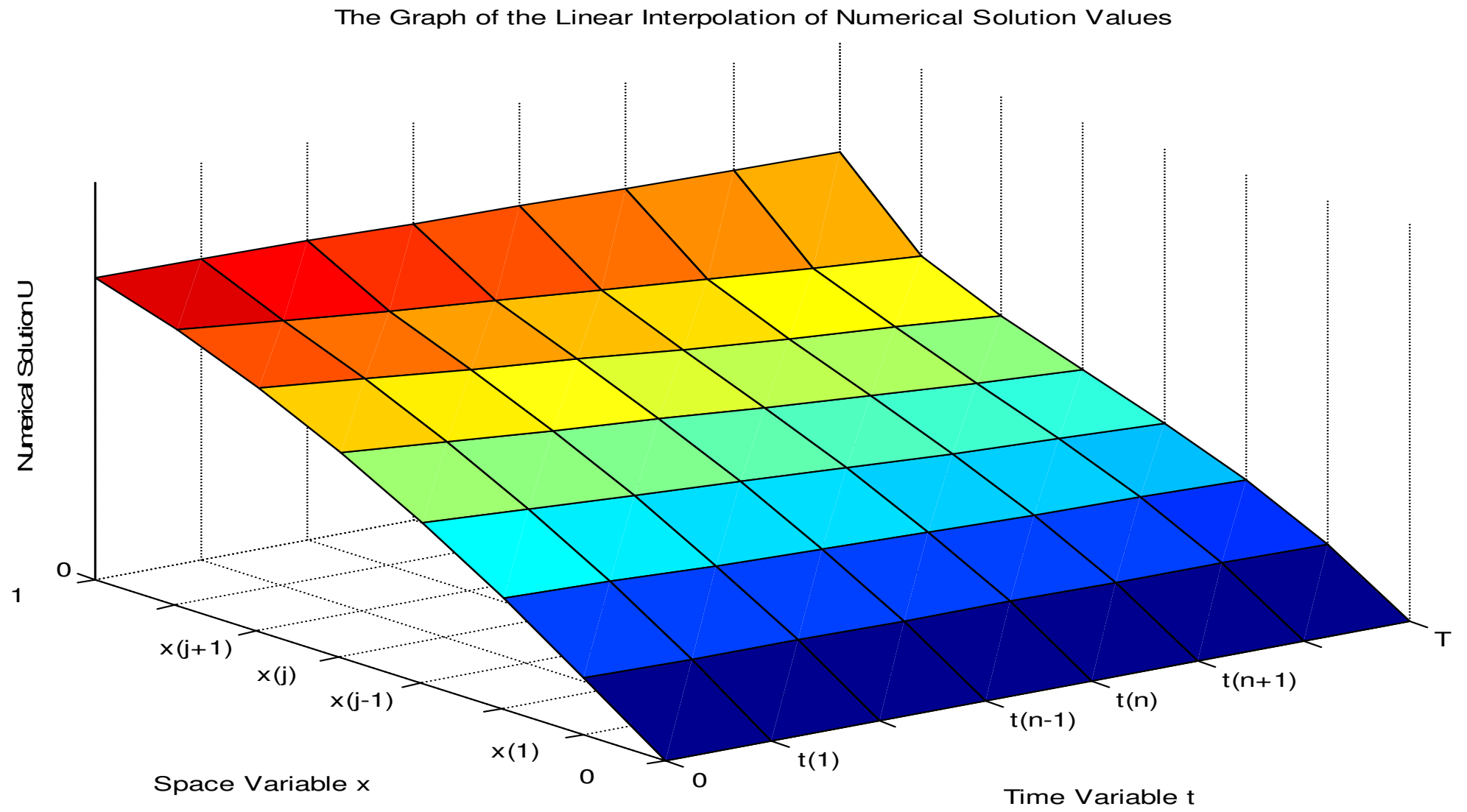

By construction, the function $\bar{s}(x,t)$ is continuous in both *x* and *t*. It also satisfies the initial condition of equation (1). It follows from (13), (27), and (41) that there is a constant $C_4'=\sqrt{2C_1(C_1+C_6)}$ such that

$$\max_j\left|U_j^n\right|\le C_4' \ , \tag{42}$$

which, in turn, implies that $\bar{s}(x,t)$ is uniformly bounded:

$$\left|\bar{s}(x,t)\right|\le C_4' \ . \tag{43}$$

Also from (27) we have: $\sum_{j=0}^{J-1}\left(\Delta_+ U_j^n\right)^2 \Delta x \le \left(C_6\right)^2$, or $\sum_{j=0}^{J-1}\left|U_{j+1}^n - U_j^n\right|^2 \le \left(C_6\right)^2 \Delta x$. This means that for each *n* and any *j*, $\left|U_{j+1}^n - U_j^n\right| \le C_6 \left(\Delta x\right)^{\frac{1}{2}}$. Or,

$$\left|\bar{s}(x+\Delta x, t) - \bar{s}(x,t)\right| \le C_6 \left(\Delta x\right)^{\frac{1}{2}}. \tag{44}$$

Also, from (34), it results that, in particular and for any *n*,

$$\left\|U^{n+1} - U^n\right\|^2 \le C_{11}\Delta t. \tag{45}$$

Now, after applying (41) to the vector $U^{n+1} - U^n$, and using (27) and (45), we obtain

$$\max_j \left|U_j^{n+1} - U_j^n\right|^2 \le \sqrt{C_{11}\Delta t}\left(4\left\|\Delta_+ U^n\right\| + 2\left\|U^n\right\|\right) \le \sqrt{C_{11}\Delta t}\left(4C_6 + 2C_1\right).$$

Therefore, there is a constant $C_5' = \sqrt{\sqrt{C_{11}}\left(4C_6 + 2C_1\right)}$ such that

$$\left|U_j^{n+1} - U_j^n\right| \le C_5'\left(\Delta t\right)^{\frac{1}{4}}. \tag{46}$$

Thus,

$$\left|\bar{s}(x, t+\Delta t) - \bar{s}(x,t)\right| \le C_5'\left(\Delta t\right)^{\frac{1}{4}} \tag{47}$$

(44) and (47) imply that $\bar{s}(x,t)$ is Hölder-continuous of order $\frac{1}{2}$ in *x* and of order $\frac{1}{4}$ in *t*. In fact, if we let $C_6' = \max\left\{C_6, C_5'\right\}$, then

$$\left|\bar{s}(x+\Delta x, t+\Delta t) - \bar{s}(x,t)\right| \le C_6'\left(\left(\Delta x\right)^{\frac{1}{2}} + \left(\Delta t\right)^{\frac{1}{4}}\right). \tag{48}$$

From (43) and (48) it follows that there is a sequence $\left\{\Delta x_i, \Delta t_i\right\}$ such that as $\left\{\Delta x_i, \Delta t_i\right\} \to 0$, the function $\bar{s}(x,t)$ converges uniformly to some functions $s(x,t)$:

$$\lim_{(\Delta x, \Delta t)\to 0}\left|\bar{s}(x,t) - s(x,t)\right| = 0. \tag{49}$$

Furthermore, from (48) and from the definitions of functions $\overline{r}(x,t)$ and $\overline{l}(x,t)$, at any point $(x,t)$ we have:

$$
\begin{aligned}
\left|\overline{s}(x,t)-\overline{r}(x,t)\right| &\le C_6'\left((\Delta x)^{\frac{1}{2}}+(\Delta t)^{\frac{1}{4}}\right),\\
\left|\overline{s}(x,t)-\overline{l}(x,t)\right| &\le C_6'\left((\Delta x)^{\frac{1}{2}}+(\Delta t)^{\frac{1}{4}}\right).
\end{aligned}
\tag{50}
$$

From (49) and (50) it follows that $\overline{r}(x,t)$ and $\overline{l}(x,t)$ also uniformly converge to $s(x,t)$ as $(\Delta x,\Delta t)\to 0$. This establishes that $r(x,t)=l(x,t)=s(x,t)$.

■

From this point on, we will use $u(x,t)$ for each of the functions $s(x,t)$, $r(x,t)$, and $l(x,t)$. It also follows immediately from lemma 2 and the previous results that $r'(x,t)=l'(x,t)$, and $r_t(x,t)=l_t(x,t)$.

We will next construct the functions $\overline{m}(x,t),\overline{m_1}(x,t),\overline{m_2}(x,t),\overline{q_1}(x,t),$ and $\overline{q_2}(x,t)$ as follows:

$$
\begin{aligned}
&\overline{m}(x,t)=\sum_{n=0}^{N-1}\sum_{j=0}^{J}\chi_{(D]_j}\frac{V_j^n}{2}, && (x,t)\in(0,1]\times(0,T],\\
&\overline{m}(x,0)=U_j^0, && x\in((j-1)\Delta x, j\Delta x],\ j=1,2,...,J,
\end{aligned}
$$

$$
\begin{aligned}
&\overline{m_1}(x,t)=\sum_{n=0}^{N-1}\sum_{j=1}^{J-1}\chi_{(D]_j}\Delta_0\left(\frac{V_j^n}{2}\right), && (x,t)\in(0,1]\times(0,T],\\
&\overline{m_1}(x,0)=\Delta_0\left(\frac{V_j^0}{2}\right), && x\in((j-1)\Delta x, j\Delta x],\ j=1,2,...,J-1.
\end{aligned}
$$

$$
\begin{aligned}
&\overline{m_2}(x,t)=\sum_{n=0}^{N-1}\sum_{j=0}^{J-1}\chi_{(D]_j}\Delta_+\left(\frac{V_j^n}{2}\right), && (x,t)\in(0,1]\times(0,T],\\
&\overline{m_2}(x,0)=\Delta_+\left(\frac{V_j^0}{2}\right), && x\in((j-1)\Delta x, j\Delta x],\ j=0,1,...,J-1.
\end{aligned}
$$

$$\overline{q_1}(x,t)=\sum_{n=0}^{N-1}\sum_{j=0}^{J-2}\chi_{(D]_j}\Delta_{++}\left(\frac{V_j^n}{2}\right), \qquad (x,t)\in(0,1]\times(0,T],$$

$$\overline{q_1}(x,0)=\Delta_{++}\left(\frac{V_j^0}{2}\right), \qquad x\in\left(j\Delta x,(j+1)\Delta x\right],\ j=0,1,...,J-2,$$

$$\Delta_{++}V_j^n=\frac{1}{2}\left(3\Delta_+V_j^n-\Delta_+V_{j+1}^n\right).$$

$$\overline{q_2}(x,t)=\sum_{n=0}^{N-1}\sum_{j=2}^{J}\chi_{(D]_j}\Delta_{--}\left(\frac{V_j^n}{2}\right), \qquad (x,t)\in(0,1]\times(0,T],$$

$$\overline{q_2}(x,0)=\Delta_{--}\left(\frac{V_j^0}{2}\right), \qquad x\in\left((j-1)\Delta x,j\Delta x\right],\ j=2,3,...,J,$$

$$\Delta_{--}V_j^n=\frac{1}{2}\left(3\Delta_-V_j^n-\Delta_-V_{j-1}^n\right).$$

Similar arguments as those about $\overline{r}$, $\overline{r}'$, $\overline{r}_t$, $\overline{l}$, $\overline{l}'$, and $\overline{l}_t$ , and in the same order of derivation, would reveal that $\overline{m},\ \overline{m_1},\ \overline{m_2}\in L_2\left([0,T];L_2\left([0,1]\right)\right)$, that they respectively converge to some functions $m,\ m_1,\ m_2$ in this space as the mesh size is shrunk, that $m_1$ and $m_2$ are the generalized derivatives of $m$ with respect to *x,* that $\overline{m}(x,t)$ also converges uniformly to the same function that $\overline{s}(x,t)$ converges to, and that $m(x,t)=s(x,t)=u(x,t)$ and $m_1(x,t)=m_2(x,t)=r'(x,t)=l'(x,t)=D_x\left(u(x,t)\right)$. Furthermore, from the definition of operators $\Delta_{++}$ and $\Delta_{--}$ it follows that for any grid-function $\{V^n\}$, $\left\|\Delta_{++}V^n\right\|\le 2\left\|\Delta_+V^n\right\|\le 2C_6$, and $\left\|\Delta_{--}V^n\right\|\le 2\left\|\Delta_-V^n\right\|\le 2C_6$, hence

$$\left\|\Delta_{++}\left(\frac{V^n}{2}\right)\right\|\le C_6,$$
$$\left\|\Delta_{--}\left(\frac{V^n}{2}\right)\right\|\le C_6. \tag{51}$$

It then follows from (51) that

$$\left\|\overline{q_1}(x,t)\right\|_E,\ \left\|\overline{q_2}(x,t)\right\|_E\le C_2'. \tag{52}$$

Inequality (52) implies that $\overline{q_1}$ and $\overline{q_2}$ converge weakly to some respective functions $q_1$ and $q_2$ in $L_2\left((0,T);L_2(0,1)\right)$.

**Lemma 3:**
$q_1(x,t) = q_2(x,t) = D_x(u(x,t))$.

**Proof:**
for any $\phi(x,t) \in \Phi$ with compact support, Let
$\bar{\phi}(x,t) = \sum_{n=1}^{N-1} \sum_{j=1}^{J-1} \chi_{(D]_j} \phi_j^n,\ (x,t) \in (0,1] \times (0,T], \bar{\phi}(x,0) = \bar{\phi}(x,T) = \bar{\phi}(0,t) = \bar{\phi}(1,t) = 0$,

so that $\lim_{(\Delta x, \Delta t) \to 0} \bar{\phi}(x,t) = \phi(x,t)$. We also note that $\lim_{\Delta x \to 0} \Delta_{++} \phi_j^n = \lim_{\Delta x \to 0} \Delta_{--} \phi_j^n = \frac{d\phi(x,t)}{dx}$.
Furthermore, since $\phi_0^n = \phi_J^n = 0$, by Taylor's expansion, $\phi_1^n,\ \phi_{J-1}^n = O(\Delta x)$. Also by (44), $V_1^n - V_0^n = O(\Delta x)^{\frac{1}{2}}$, and $V_J^n - V_{J-1}^n = O(\Delta x)^{\frac{1}{2}}$. Therefore, we have:

$$\langle q_2(x,t), \phi(x,t) \rangle_E = \int_0^T \int_0^1 \left( \lim_{(\Delta x, \Delta t) \to 0} \sum_{n=0}^{N-1} \sum_{j=2}^{J} \chi_{(D]_j} \Delta_{--} \frac{V_j^n}{2} \right) \left( \lim_{(\Delta x, \Delta t) \to 0} \sum_{n=1}^{N-1} \sum_{j=1}^{J-1} \chi_{(D]_j} \phi_j^n \right) dx\, dt$$

$$= \frac{1}{2} \lim_{(\Delta x, \Delta t) \to 0} \int_0^T \int_0^1 \sum_{n=0}^{N-1} \left( \sum_{j=2}^{J} \frac{1}{2\Delta x} \left( 3V_j^n - 4V_{j-1}^n + V_{j-2}^n \right) \phi_{j-1}^n \right) dx\, dt$$

$$= - \lim_{(\Delta x, \Delta t) \to 0} \int_0^T \int_0^1 \sum_{n=0}^{N-1} \left( \left( \sum_{j=0}^{J-2} \left( \frac{V_{j+1}^n}{2} \right) \left( \lim_{\Delta x \to 0} \frac{-3\phi_j^n + 4\phi_{j+1}^n - \phi_{j+2}^n}{2\Delta x} \right) \right) + \frac{O(\Delta x)^{\frac{3}{2}}}{2\Delta x} \right) dxdt$$

$$= - \lim_{(\Delta x, \Delta t) \to 0} \left\langle \frac{d}{dx} \phi(x,t), \bar{m}(x,t) \right\rangle_E + \lim_{\Delta x \to 0} \left( O(\Delta x) \right) = - \left\langle \frac{d}{dx} \phi(x,t), u(x,t) \right\rangle_E .$$

■

We will now state and prove the main result:

**Theorem 1:**
Let $D = (0,T) \times (0,1), T > 0$, be a domain, in where, $A(x,t)$ is differentiable in *t* and twice differentiable in *x*, $B(x,t)$ is continuously bounded in *t* and differentiable in *x*, $C(x,t)$ is continuous in both *x* and *t*, and $f(x) \in L_2([0,1])$. Consider the degenerate parabolic equation (4)-(7) along with one of the four boundary characteristics (8.a), (8.b), (8.c), and (8.d). Then there exists a unique solution in $W_2^1((0,T); W_2^1(0,1))$ if

(i) $B(x,t)$ has characteristic (8.a).

(ii) $B(x,t)$ has characteristic (8.b) with $B(0,t) \geq \frac{1}{7} \frac{d}{dx} A(0,t)$ and $B(1,t) + \frac{d}{dx} A(1,t) \geq 0$, and a boundary condition is given at $x = 1$.

(iii) $B(x,t)$ has characteristic (8.c) with $B(1,t) \le \frac{1}{7}\frac{d}{dx}A(1,t)$ and $B(0,t) + \frac{d}{dx}A(0,t) \le 0$,

and a boundary condition is given at $x = 0$.

(iv) $B(x,t)$ has characteristic (8.d) and boundary conditions are given at $x = 1$ and $x = 0$.

(Note: The requirements that boundary conditions be given at $x = 1$ or at $x = 0$ or at both, are meant to satisfy the uniqueness property.)

**Proof:**

We will prove the existence and uniqueness for case (ii) as indicated in the theorem. Case (iii) is similar, and cases (i) and (iv) require less work.

By integration by parts, we must show that for any test function $\phi(x,t) \in \Phi$ with compact support, the function $u(x,t)$ satisfies:

$$\left\langle D_t\, u(x,t), \phi(x,t) \right\rangle_E = -\left\langle A(x,t)\frac{d\,\phi(x,t)}{dx}, D_x\, u(x,t) \right\rangle_E + \left\langle B(x,t)\frac{d\,\phi(x,t)}{dx}, D_x\, u(x,t) \right\rangle_E . \tag{53}$$

We multiply both sides of scheme (10)-(12) by $\phi_j^{n+1}\Delta x\,\Delta t$, and sum over *j* and *n,* $j = 0,1,\dots,J$; $n = 0,1,\dots N-1$, and take limit as $(\Delta x, \Delta t) \to 0$ to obtain

$$\lim_{(\Delta x,\Delta t)\to 0}\left( \sum_{n=0}^{N-1}\sum_{j=0}^{J}\left(\Delta_t U_j^n\right)\phi_j^n \Delta x \Delta t \right) = \lim_{(\Delta x,\Delta t)\to 0}\left( \sum_{n=0}^{N-1}\left(H_1 + H_2\right)\Delta t \right),$$

where,

$$H_1 = \frac{1}{2}\sum_{j=1}^{J-1}\Delta_+\left(A_{j-1/2}^{n+1/2}\left(\Delta_- V_j^n\right)\right)\phi_j^{n+1}\Delta x .$$

$$\begin{aligned} H_2 = {} & \sum_{j=1}^{J-1}\theta_j\, B_j^{n+1/2}\left(\Delta_0 \frac{V_j^n}{2}\right)\phi_j^{n+1}\Delta x + \sum_{j=1}^{i-1}\left(1-\theta_j\right)B_j^{n+1/2}\left(\Delta_{++}\frac{V_j^n}{2}\right)\phi_j^{n+1}\Delta x \\ & + \sum_{j=1}^{i-1}\left(1-\theta_j\right)B_j^{n+1/2}\left(\Delta_{--}\frac{V_j^n}{2}\right)\phi_j^{n+1}\Delta x + \sum_{j=k}^{J-1}\left(1-\theta_j\right)B_j^{n+1/2}\left(\Delta_{++}\frac{V_j^n}{2}\right)\phi_j^{n+1}\Delta x. \end{aligned}$$

Since $A_{1/2}^{n+1/2} = O(\Delta x)$, $A_{J-1/2}^{n+1/2} = O(\Delta x)$, $V_1^n - V_0^n = O(\Delta x)^{\frac{1}{2}}$, and $V_J^n - V_{J-1}^n = O(\Delta x)^{\frac{1}{2}}$, by summation by parts, we obtain:

$$\lim_{(\Delta x,\Delta t)\to 0}\left(\sum_{n=0}^{N-1}\sum_{j=0}^{J}\Delta_t U_j^n \phi_j^n \Delta x\Delta t\right) = -\lim_{(\Delta x,\Delta t)\to 0}\sum_{n=0}^{N-1}\left(\sum_{j=1}^{J-1} A_{j+1/2}^{n+1/2}\left(\Delta_+ \frac{V_j^n}{2}\right)\left(\Delta_+\phi_j^{n+1}\right) + O\left(\Delta x\right)^{\frac{1}{2}}\right)\Delta x\Delta t$$

$$+\lim_{(\Delta x,\Delta t)\to 0}\sum_{n=0}^{N-1}\left(\sum_{j=1}^{J-1}\theta_j B_j^{n+1/2}\left(\Delta_0\frac{V_j^n}{2}\right)\phi_j^{n+1}\right)\Delta x\Delta t$$

$$+\lim_{(\Delta x,\Delta t)\to 0}\sum_{n=0}^{N-1}\left(\sum_{j=1}^{i}\left(1-\theta_j\right)B_j^{n+1/2}\left(\Delta_{++}\frac{V_j^n}{2}\right)\phi_j^{n+1}\right)\Delta x\Delta t$$

$$+\lim_{(\Delta x,\Delta t)\to 0}\sum_{n=0}^{N-1}\left(\sum_{j=i}^{k}\left(1-\theta_j\right)B_j^{n+1/2}\left(\Delta_{--}\frac{V_j^n}{2}\right)\phi_j^{n+1}\right)\Delta x\Delta t$$

$$+\lim_{(\Delta x,\Delta t)\to 0}\sum_{n=0}^{N-1}\left(\sum_{j=k}^{J-1}\left(1-\theta_j\right)B_j^{n+1/2}\left(\Delta_{++}\frac{V_j^n}{2}\right)\phi_j^{n+1}\right)\Delta x\Delta t + \lim_{(\Delta x,\Delta t)\to 0}\left(O\left(\Delta x\right)\right).$$

Hence,

$$\lim_{(\Delta x,\Delta t)\to 0}\left(\int_0^T\int_0^1 \overline{r_t}(x,t)\,\phi(x,t)\,dx\,dt\right) = -\lim_{(\Delta x,\Delta t)\to 0}\left(\int_0^T\int_0^1 A(x,t)\,\overline{m_2}(x,t)\left(\lim_{\Delta x\to 0}\frac{\phi_{j+1}^{n+1}-\phi_j^{n+1}}{\Delta x}\right)dx\,dt\right)$$

$$+\lim_{(\Delta x,\Delta t)\to 0}\left(\int_0^T\int_0^1\theta(x)\,B(x,t)\,\overline{m_1}(x,t)\phi(x,t)dx\,dt\right)$$

$$+\lim_{(\Delta x,\Delta t)\to 0}\left(\int_0^T\int_0^{i\Delta x}\left(1-\theta(x)\right)B(x,t)\,\overline{q_1}(x,t)\phi(x,t)dx\,dt\right)$$

$$+\lim_{(\Delta x,\Delta t)\to 0}\left(\int_0^T\int_{i\Delta x}^{k\Delta x}\left(1-\theta(x)\right)B(x,t)\,\overline{q_2}(x,t)\phi(x,t)dx\,dt\right)$$

$$+\lim_{(\Delta x,\Delta t)\to 0}\left(\int_0^T\int_{k\Delta x}^{1}\left(1-\theta(x)\right)B(x,t)\,\overline{q_1}(x,t)\phi(x,t)dx\,dt\right).$$

Therefore,

$$\lim_{(\Delta x,\Delta t)\to 0}\left\langle\overline{r_t}(x,t),\phi(x,t)\right\rangle_E = -\lim_{(\Delta x,\Delta t)\to 0}\left\langle A(x,t)\frac{d\,\phi(x,t)}{dx},\overline{m_2}(x,t)\right\rangle_E$$

$$+\lim_{(\Delta x,\Delta t)\to 0}\left\langle\theta(x)B(x,t)\frac{d\,\phi(x,t)}{dx},\overline{m_2}(x,t)\right\rangle_E$$

$$+\lim_{(\Delta x,\Delta t)\to 0}\left\langle\left(1-\theta(x)\right)B(x,t)\frac{d\,\phi(x,t)}{dx},\overline{m_2}(x,t)\right\rangle_E .$$

Or,

$$\left\langle D_t\,u(x,t),\phi(x,t)\right\rangle_E = -\left\langle A(x,t)\frac{d\,\phi(x,t)}{dx},D_x\,u(x,t)\right\rangle_E + \left\langle B(x,t)\frac{d\,\phi(x,t)}{dx},D_x\,u(x,t)\right\rangle_E .$$

■